\documentclass[]{article} 

\usepackage[utf8]{inputenc}
\usepackage[T1]{fontenc}
\usepackage{graphicx}
\usepackage{tabularx}
\usepackage{hyperref}

\title{An Approach for Discovering Traceability Links between Regulatory Documents and Source Code Through User-Interface Labels}
\author{Antoine Mischler\\DooApp \and Martin Monperrus\\University of Lille \& Inria}
\date{}
\begin{document}
\maketitle 

\begin{abstract}
In application domains that are regulated, software vendors must maintain traceability links between the regulatory items and the code base implementing them.
In this paper, we present a traceability approach based on the intuition that the regulatory documents and the user-interface of the corresponding software applications are very close.
First, they use the same terminology.
Second, most important regulatory  pieces of information appear in the graphical user-interface because the end-users in those application domains care about the regulation (by construction).
We evaluate our approach in the domain of green building.
The evaluation involves a domain expert, lead architect of a commercial product within this area.
The evaluation shows that the recovered traceability links are accurate.
\end{abstract}

\section{Introduction}

Some companies develop software applications in domains that are regulated. For instance, in the USA, healthcare related software products  must comply with the Health Insurance Portability and Accountability Act (HIPAA).

Regulation, as software, evolves and it is essential to have traceability links between the regulation artifacts and the code artifacts \cite{sannier:tel-00941881}.
Then, to modify a software application in respond to regulatory changes, software development teams follow those traceability links and are quickly pointed to the software modules to change.
The whole point of research on traceability is on how to set up and maintain those links accurately.

The contribution of this paper is a novel fully-automated approach for recovering traceability links between regulatory documents and source code. This approach is based on the intuition that the labels of graphical user interfaces (GUIs) can act as proxy to the domain-specific terminology. If a user-interface label uses specific terms of requirements of regulatory documents, it is very likely that the code elements interacting with the label are indeed related to the requirements under consideration.

Our approach takes as input the code base of a software application that implements a specific regulation. This application must contain a human-machine interface. An indexing phase stores all user interface labels. When a developer wants to locate code elements that relate to a piece of regulation (whether a header, a sentence or a paragraph), she/he gives it as query to the system, which retrieves back a list of code elements (e.g. a list of classes in object-oriented software). The matching is done between the regulation terminology and the user-interface terminology.

The added value of our approach is the productivity gain when developers want to modify or add new features in a code base implementing a regulation that is unknown to them. Our traceability system directly points them to relevant pieces of code. To this extent, the goal is not to absolutely retrieve all related code elements but to retrieve quickly good ones. Once one element is found, the collaborating classes are probably also of interest. 

Logically, we evaluate our approach as a recommendation system.
First, we extracted 53 requirements from the official documents regulating the construction of green buildings in Europe.
Then, we run our traceability system on those 53 requirements and the code base of a proprietary product complying with the regulation.
The first author, expert of the code base, assessed the relevance of the top 25 recovered traceability links.
Overall, for 44/53 queries, our prototype system found at least one correct traceability link. 
For 19/44 answered queries, the first recommended code item was correct. 

To sum up, our contributions are:
\begin{itemize}
\item a fully-automated approach for recovering traceability links between regulatory documents and source code based on user interface labels;
\item an evaluation on a real work regulation and a complying  commercial product, done by a domain expert.
\end{itemize}

The remainder of this paper is as follows.
Section \ref{fig:approach} describes our approach,
section \ref{sec:evaluation} presents its evaluation,
section \ref{sec:rw} discusses the related work and section \label{sec:conclusion} suggests future work.

\section{Approach}
\label{sec:approach}

\begin{figure}
\includegraphics[width=\columnwidth]{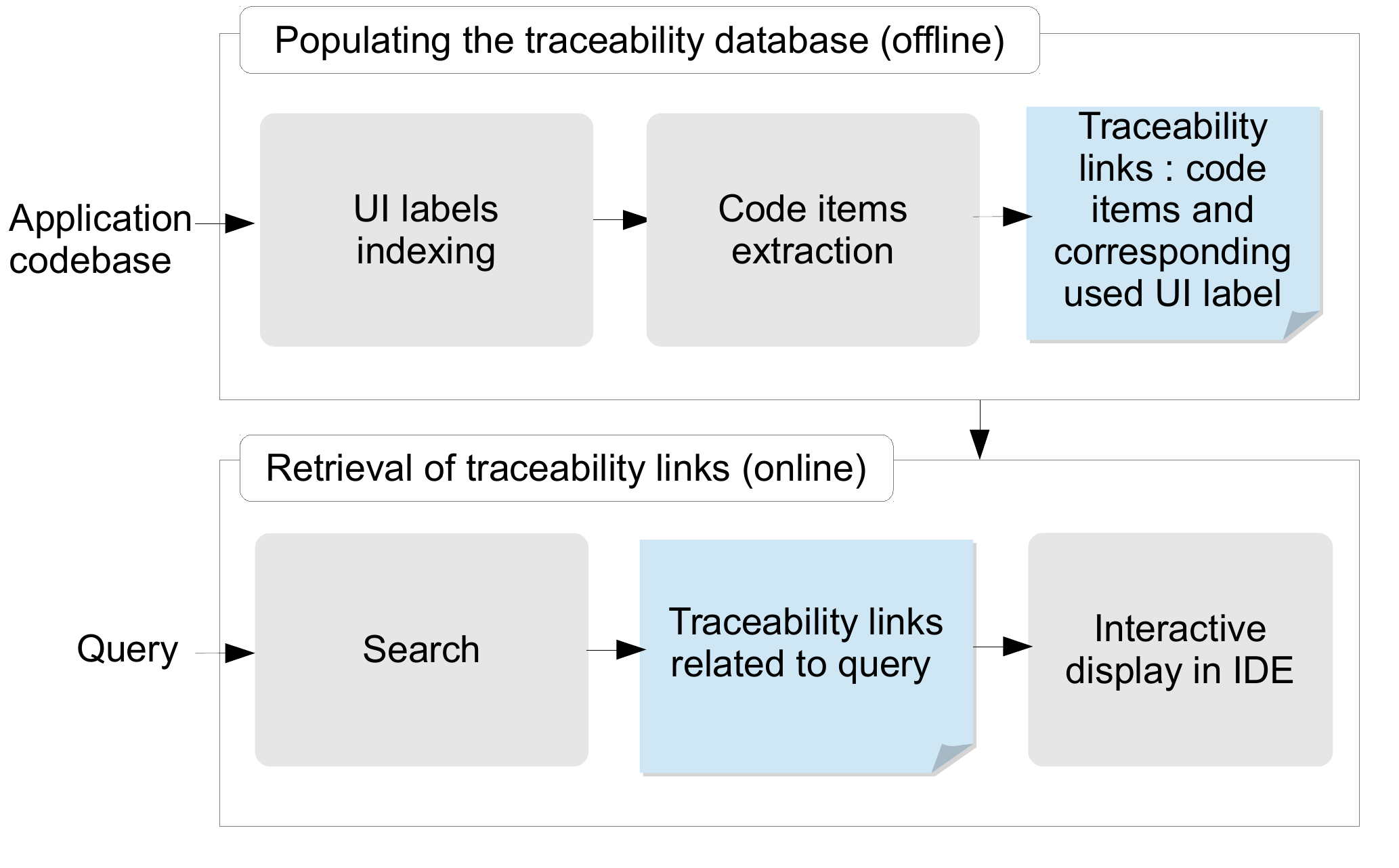}
\caption{Traceability link extraction overview.}
\label{fig:approach}
\end{figure}

Our traceability approach is based on the intuition that most regulation items will result, at least partially, in a label in the user-interface of the application. The whole process has two steps: first, we populate a traceability database by analyzing user-interface labels. Then, we automatically extract the traceability links by querying this database to retrieve the related code items. This process is summarized in Figure \ref{fig:approach}.

Our approach is both generic and specific. It is generic in the sense that it can be applied to any application domain. 
It is specific in the sense that the tool implementing the norm must have a user-interface. Even if the user-interface only spans a subset of features or requirements of the regulation documents, the traceability provided in this scope it is valuable.

\subsection{Populating the traceability database offline}
\label{sec:populating_database}

\paragraph{UI labels extraction}
The first step consists in extracting all user interface labels from the application source. In large internationalized applications, UI labels are generally stored in resource files. A resource file is a map of key-value pairs for a given language. Each entry in the resource bundle represents the translation of an UI label for this language. The value is the translation of the key and the key is used in the code to reference this label. At the end of this step, we obtain a database with all labels used in the UI.

\paragraph{Code items extraction}
\label{sec:code-extraction}
The second step consists in retrieving the code items that use the labels collected in the previous step. In our database, each label is paired with a key that is used in the code to reference that particular label. In most cases, a literal search of the key in the codebase allows us to retrieve the code item related to the UI label. 

In some cases, no result matches the key of the UI label. There two reasons for this.
First, this happens when the label is not used in the UI, it is either deprecated or unused due to an accidental label duplication after a refactoring.
Deprecation of labels happens during the lifecycle of an application, when developers update the UI and forget to clean the deprecated resources. 

The second reason is that the key is sometimes dynamically built at runtime and can not be retrieved using a literal search in the codebase. In this case, we propose an approach based on dynamic traces.
This involves generating a database of runtime traces of the application. The application is used according to a usage scenario and a monitor logs all  accesses to the UI label resources. 
This enables us to recover links between UI labels and code elements that can not be observed statically.
This approach has a limitation: we cannot guarantee that all dynamic keys have been used during the runtime scenario.  

\paragraph{Synonym Normalization}
Synonym normalization \cite{leuser2010} consists in replacing synonymous terms for a single concept by the same word. In the domain of air tightness, the regulation defines a list of terms and symbols that might be used as synonyms for frequently used values or concepts. For example, ``n50'' can be used to represent the ``taux de renouvellement d'air sous 50 Pa''. Our prototype system implements synonym normalization as preprocessing to indexing. Hence, searching for ``n50'' of its long explicit version yields the same results. 

\subsection{Online retrieval of traceability links}
\label{sec:online_retrieval}
The database of traceability links populated offline is then used to retrieve traceability links on demand in an online manner.

\paragraph{Search}
To find code elements related to a regulatory item, a developer types a query in natural language. 
For instance, in the domains of cash machines, the developer might type ``incorrect password''. The traceability system then answers that class ``ScreenA'' and ``ScreenB'' uses an UI label that matches the query ``incorrect password''.

Given the developer query (e.g. a regulatory item), we select relevant UI labels from the database using an information retrieval library. A tf-idf (term frequency-inverse document frequency\footnote{see \url{https://en.wikipedia.org/wiki/Tf-idf}, last accessed March 11 2014}) statistic is used to find the labels related that are most related to the query. 

For each matched label, the source code classes that use this label are retrieved using the literal or trace search described in Section \ref{sec:code-extraction}.
If a matching label is not used according to both techniques, the developer is told that a matching label has been found but no code elements can be identified. It is her responsibility to tell whether the label is a dead one or not. From the traceability viewpoint, one can not be state beforehand that the \{key,label\} is actually not used in the application.

To sum up, a search results in a collection of traceability links matching the user query.

\paragraph{Integrated display in IDE}
An IDE integration of the search process is important for an easy use of the traceability system during daily development. The IDE integration should allow the user to enter its query and displays the corresponding traceability links. Our prototype is presented in section \ref{sec:prototype_implementation}.

\subsection{Prototype implementation}
\label{sec:prototype_implementation}
We have developed a prototype that implements our approach. Our prototype consists of two artifacts.
First,  a Java library performs the traceability population (see \ref{sec:populating_database}) and the online retrieval of traceability links  (see \ref{sec:online_retrieval}).
Second,  the integration of the approach into an IDE is demonstrated with a plugin for the Intellij IDEA development environment.

\paragraph{Library}
The traceability database if first populated by scanning the applications resources and retrieving all resource bundles. Our library indexes key-value pairs using Apache Lucene \cite{LUCENE}, a well known and mature indexing open-source application written in Java. 
The indexing uses the language specific indexing components of Lucene (e.g. \texttt{FrenchAnalyzer} if the regulation and user-interface are in French). Those components perform tokenization, stop word removal and stemming in a language specific manner.

Then, we analyze all source files of the application to extract the resource bundles keys used in the graphical user-interface. Our implementation uses Spoon \cite{SPOON} to analyze the Java source code and extract all literals. These literals are then stored in memory to keep track of their usage in source code.

\paragraph{Intellij IDEA plugin}
The traceability UI is implemented as an Intellij IDEA plugin. This plugin provides a menu entry that allows the user to enter its search query. The library is then used to perform this online query and returns the collection of traceability links. These links are displayed in the search result window of the IDE. For each link, the plugin displays the corresponding resource bundle key-value and the weight of the result. By clicking on a link, the IDE opens the Java class where the key usage has been identified.

The plugin is available in the JetBains Plugin Repository and can be downloaded and tested \cite{NLCSPLUGIN}.

\section{Evaluation}
\label{sec:evaluation}

We now evaluate our approach to recover traceability links between source code and regulatory documents.
We set up an evaluation process in the context of the first author's company: green building.

\subsection{Background}

\paragraph{Application Domain}
DooApp develops and sells a product for measuring and reporting on the thermal performance of buildings with respect to air tightness.
The product, called Infiltrea, drives a blower door and generates a report that is part of the certification of green buildings.

The first author of this paper is the co-founder of a company called DooApp, that develops and sells a product for measuring and reporting on the thermal performance of buildings with respect to air tightness.
The product, called Infiltrea, drives a blower door and generates a report that is part of the official certification of green buildings.

\paragraph{Regulations}
The measure of air permeability is regulated by a set of regulatory documents at the European and French level, in particular EN-13829 \cite{AFNOR2001}.
The regulation allows one to compare test results from different test operators and countries.
The reports generated by Dooapp's product Infiltrea, complies with this regulation.

\begin{figure*}
\includegraphics[width=\textwidth]{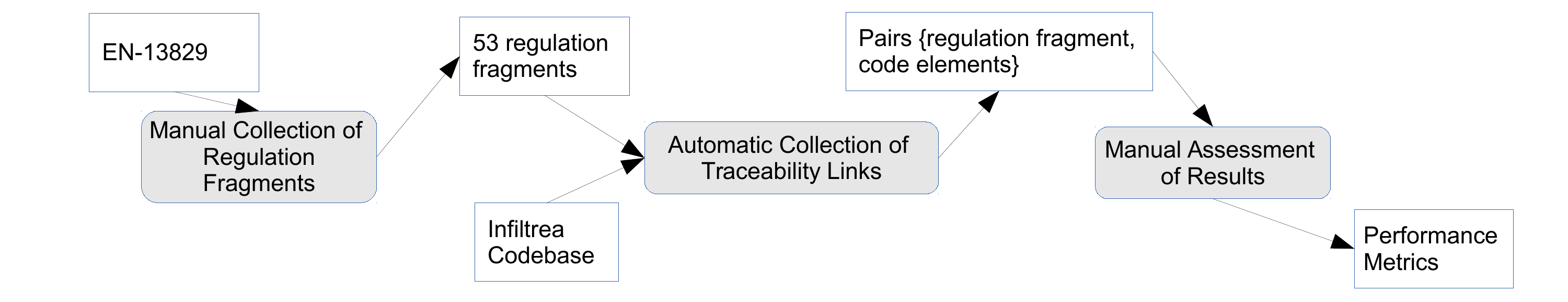}
\caption{Our Evaluation Process. The collection of regulation fragments and the manual assessment of results is done by an expert who masters both the application domain (green building) and code base of the application under study.}
\label{evaluation-process.pdf}
\end{figure*}

\subsection{Evaluation Process}

The evaluation is as follows.
First, we have collected 53 regulation fragments in EN-13829 \cite{AFNOR2001}.
Each of those regulation fragments is an atomic feature of the thermal measurement product under study.
Second, for each of them, we have used our prototype implementation to extract the matching pieces of code in Infiltrea's codebase.
Third, the first author (the domain and codebase expert) has assessed each of the top-20 matched source code elements for each of the 53 regulatory items to see whether they were correct or not.

In this context, there are two potential issues: false positive and false negative answers.
False positive answers are code elements that our system points as related to the query while they actually have nothing to do with the regulation item under study.
False negative answers are related code elements that are not identified as such during traceability recovery.
Since our approach is meant to be used as a recommendation system to gain productivity, having some false negatives is not an issue\footnote{The verification process later in the process (out of scope of our approach) is responsible for checking that nothing has been forgotten during the implementation of the change.}. 
Each false positive is wasted time for the developer. Consequently, we try to minimize the false positives and to some extent the false negatives. 
To capture those two facets, we compute performance metrics such as precision and recall.

\subsection{Dataset Setup}

Within EN-13829, we have identified 53 key regulatory items to be used as evaluation dataset.
For instance, a regulation item states that the measurement process involves ``Computing the air flow rate at reference pressure''.
In this expression, the key words are  ``reference pressure'' and ``air flow rate'': a software application in the domain of EN-13829 must handle the notion of reference pressure. So, we can except these key words to be used in the graphical user-interface.

Let us now consider a developer who wants to find the code implementing this regulatory item. One natural solution is to use the regulatory sentence as ``input query'' of a traceability system. This is the reason for which the 53 regulatory items form the evaluation dataset of our system.

Each regulatory item is saved in two forms: a short form representing the key description sentence such as ``Computing the air flow rate at reference pressure'', and a long form composed of the full paragraph extracted from the regulation. 

Since the first author is an expert of the application domain, we think that the 53 regulatory items are valid both in terms of importance and scope.
By importance, we mean that those regulation fragments indeed matters from a business value viewpoint.
By scope, we mean that the identified fragments cover one and only one domain feature or constraint.

The short regulation items have an average length of 9 words and the long form an average length of 44 words.

\subsection{Traceability Link Collection}

For each of the 53 regulation items them, we have used our prototype implementation described in Section \ref{sec:approach} to extract the related pieces of code in the code base of Infiltrea (the software product developed by the company of the first author).

\begin{figure}
\includegraphics[width=\columnwidth]{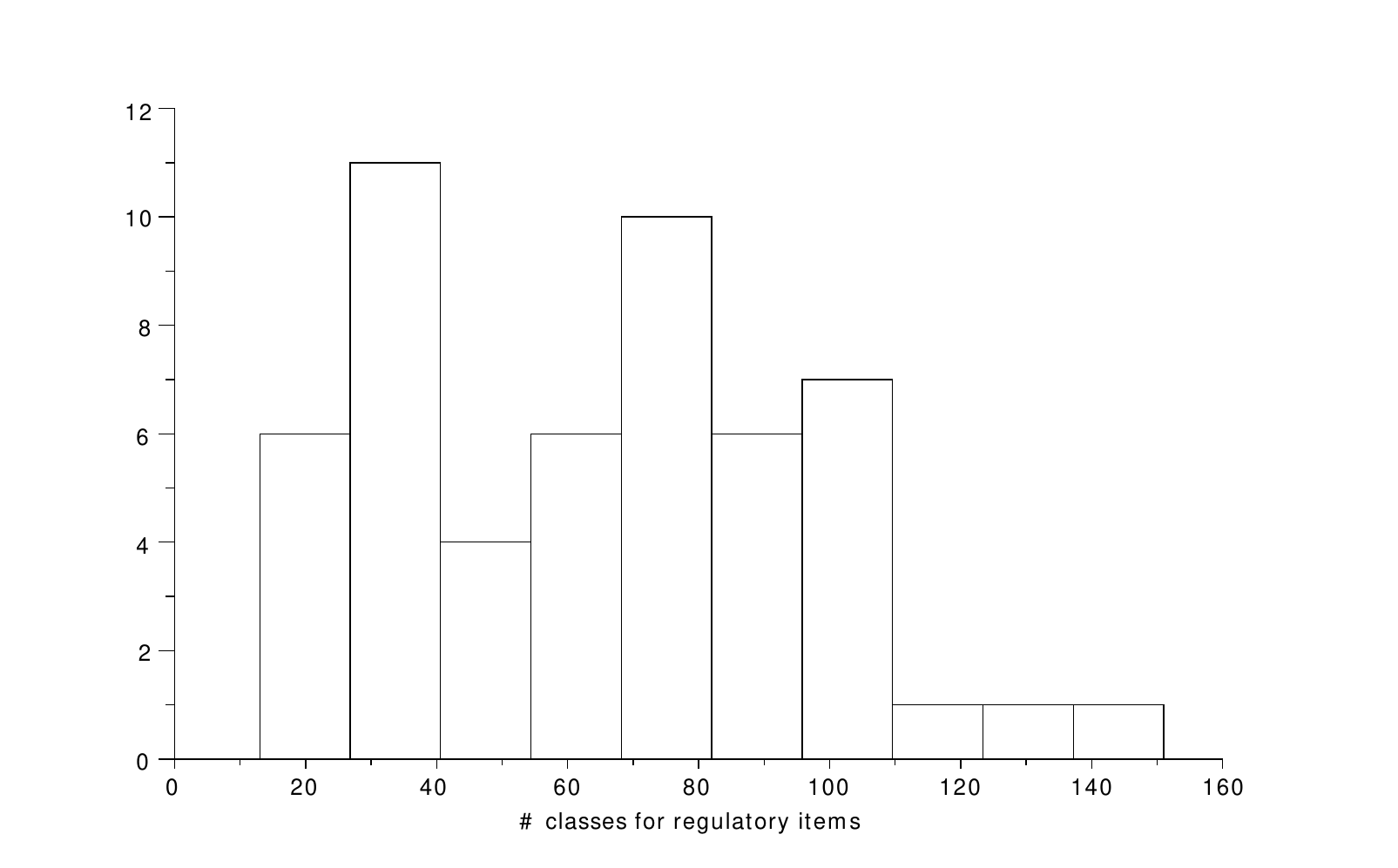}
\caption{Distribution of the number of traceability results for the 53 considered regulatory items. The median number of referred Java classes is 65; showing that many UI labels match more or less closely the regulation items.}
\label{fig:distrib}
\end{figure}

Figure \ref{fig:distrib} gives the distribution of the number of traceability results for the 53 considered regulatory items (number of user interface key-value pairs). 
A traceability link consists of a pair \{user-interface label, code element\}.
The median number of traceability links is 65; showing that many UI labels match the regulation items (more or less closely).
\emph{This is a first piece of evidence that our intuition holds: the UI labels indeed capture the terminology and spirit of the regulation}.

\subsection{Traceability Link Assessment}
\label{sec:assessment}

\begin{figure}
\centering\includegraphics[width=7.3cm]{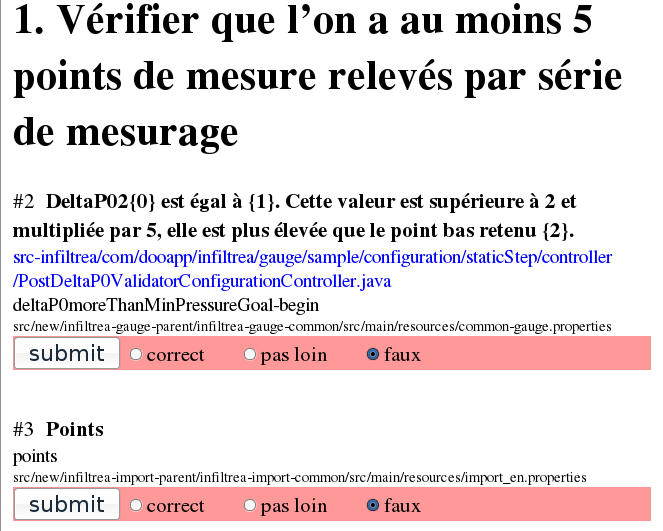}
\caption{Screenshot of the web-based assessment tool (in French). The domain and codebase expert rates on a Likert scale whether the traceability link is appropriate: correct (``correct''), related (``pas loin''), wrong (``faux'').}
\label{fig:screenshot-evaluation-tool.png}
\end{figure}

For each regulatory item of the dataset, the first author manually assessed up to 20 traceability links (in the top-20) as given by our prototype implementation.
He assessed the results for both the short and long version of the regulatory items.
The assessment was supported by a specific web-application tool shown in Figure \ref{fig:screenshot-evaluation-tool.png}.
In this evaluation tool, the traceability links are presented with four components:
the UI label (for instance ``DeltaP02 est égal \ldots''),
the Java file that uses this UI label (in blue),
the key used to refer to this label (the key name usually has some semantic meaning), 
and the resource file name defining the pair \{key,label\}.
The two first components are the essential part of the traceability link (the label and the corresponding Java class).

The last two components are specific to the UI technology we are considering. 
Still, it is important to present them to the domain and code expert because it helps him to quickly grasp which part of the code is involved in the traceability link, better than with only the class name.

For each traceability link, the domain and codebase expert rates on a Likert scale whether the traceability link is appropriate: correct, related (``pas loin'' in French), and wrong (``faux'' in French).
``Correct'' means that the assessed traceability link refers to a class that actually implements the regulatory item.
``Related'' means that it refers to a class that does not directly implement the regulatory item, but closely collaborate with a class that implements it.
``Wrong'' means that it refers to a class that does not implement the regulatory item under consideration.

In total, we have collected 1473 manual assessment of retrieved traceability links. This evaluation data is then used as input to compute performance metrics.

\subsection{Metrics}

Our traceability system is a recommendation system. 
Developers query it with a regulation related query (a sentence or a paragraph) and in response, the system recommends a list of code elements that are very probably related to the query (a list classes of object-oriented systems in our prototype system) 
To evaluate it, we use precision and recall.
The golden truth is the manual assessment data presented in the previous section \ref{sec:assessment}.
Precision and recall are computed at the top 5 and top 10 positions.

Table \ref{tab:evaluation} gives the results of this evaluation.
The traceability system is queried with both the short version of the regulatory items (one sentence) and the long version (one paragraph).

Let us consider the performance for the short queries (middle column).
Over the 53 real world queries, the system finds a correct traceability link for 44 of them.
For 19/44 answered queries, the first recommended code item is correct, it actually implements the requirement that is given as input (this may not be the only related code elements: in this context, ``correct'' does not mean complete).
Overall, the median rank of the first correct traceability link is 2.
This means that for most queries, the developer finds a correct traceability link in the top 2 results.
Over the top 5 returned results, the average precision is 0.22, which means that in average one code elements out of the top five recommended ones is relevant to the query.
This number means that there is almost a good traceability link in the top 5 results (given that there are often a single code element to find, in this case the precision is $1/5=0.2$ by construction).

For the long version (second column of Table \ref{tab:evaluation}), the precision and recall decrease for the top 5 and top 25 results.
This corresponds to the fact that the long version contains a lot of words that interfere with the core of the query.
This noise decreases the quality of the results. However, one query that was unanswered with the short version becomes answered with the long version. This is expected, the long version contains more word that can match UI labels.

This evaluation based on real regulations and manual expert assessment shows that our UI-based approach to traceability actually works in practice.
For 44/53 queries, the system finds at least one correct traceability link. 
For 19/44 answered queries, the first recommended code item is correct. 
There are good reasons to trust the results, since the first author is a domain and codebase expert:
1) the 53 evaluation query extracted from the regulatory documents are realistic and important in the considered application domain (domain expertise);
2) the 1473 codebase elements pointed by the traceability links have been evaluated by a codebase expert to rate the links (code expertise).

\begin{table}[t]
\begin{tabularx}{\textwidth}{X|p{2cm}|p{2cm}}
\textbf{Metric} & \textbf{Short query}& \textbf{Long query}\\
\hline
Number of regulatory items & 53 & 53 \\
Number of queries & 53 & 53\\

Answered queries & 44/53 (83\%)& 45/53 (85\%)\\

Median rank of first correct traceability link & 2 & 5\\

\# queries where result \#1 is correct & 19/44 (43\%)  &  16/44 (36\%) \\

Average precision (top 5)& 0.22 & 0.14 \\
Average recall (top 5)& 0.46 & 0.28 \\

Average precision (top 20)& 0.10 & 0.07 \\
Average recall (top 20)& 0.68 & 0.48 \\

\hline
\end{tabularx}
\caption{The evaluation metric of our traceability system. Over 53 real world queries, the system is capable of correctly answering most of them. When it answers, the first correct traceability link is very often in the top results.}
\label{tab:evaluation}
\end{table}

\subsection{User Study}
\label{sec:user_study}
Finally, we performed a small user study of the tool.
Within a weekly 1-hour technical seminar, we presented the traceability tool 
to the 3 developers of Infiltrea. We then asked them to use it in an informal manner, related to their current development duties.
Those 3 developers are the exact type of users targeted by our traceability approach: they work in a regulated domain and are daily performing maintenance tasks on a large code base implementing the corresponding regulation. They do not master the whole code and regularly need help to locate relevant code elements.

After 1 hour of free usage, the feedback was positive (note there is a clear bias due to the fact that the new IDE feature was proposed by a colleague).
A good anecdote is the following:
the only case when a developer complained, he realized that he looked for a feature he was about to develop. Since the feature was not yet implemented hence no traceability was found. The query was meaningful, it was related to the regulation, but no traceability link could ever be found by construction.

This small user study gives us confidence that the tool could be valuable for the developers. 
However, beyond the bias towards positive feedback, the developers were novice neither in the domain nor in the code base.
A more comprehensive and expensive user study would simulate this situation.
Today, the tool is still in use in the development team of the company.

\subsection{Limitations}
\label{sec:limitations}

The main limitation of our approach is that it only works with graphical software applications (whether desktop or web-based). For other types of software, we need another proxy to the domain terminology, such as the code element names (as usually done, see our discussion on related work e.g. \ref{sec:rw}).

Our approach is based on the terminological matching between the regulation in natural language and the user-interface labels (also in natural language). It is not tied to a particular language. If both the regulation and the user-interface are written in Tamil for instance, it would work. However, there is one case where our approach is not applicable: when the regulation is in one language (say English) and the software application implementing it in another one (say French).

Within graphical software applications, we have studied only one application domain (air permeability measurement). It may be the case that within this application domain, the terminology and concepts of the regulatory documents is more present in the user-interface than in other domains. If true, this is a case to the external validity of our empirical results.

\section{Related Work}
\label{sec:rw}

Winkler and Pilgrim \cite{winkler2010survey} and Spanoudakis and Zisman \cite{spanoudakis2005software} provide an introduction and overview to software requirements traceability. They describe the different types of traceability relations and present the existing approaches to traceability mining including manual, semi-automatic and automatic approaches. On the one side, manual approaches require the developer or analyst to track each link by hand. On the other side, automatic approaches involve generating links between the requirement and the code artifacts dynamically.

Antoniol et al. \cite{antoniol2002recovering} introduced automatic traceability based on Information Retrieval (IR) methods. They compared a Vector Space Model and a probabilistic approach \cite{antoniol2002recovering}. Both methods produced promising results. Marcus and Maletic \cite{marcus2003recovering} explored Latent Semantic Indexing methods  obtaining results as good as Antoniol et al. with less preprocessing. The effectiveness of IR techniques in traceability have been furthered examined by De Lucia \cite{lucia2007recovering} and Hayes et al. \cite{hayes2006advancing}.

One issue encountered by the previous methods relates to terminology mismatch. Code artifacts do not always use the same terminology as in the requirements. Cleland-Huang et al. \cite{cleland2010machine} focused on machine learning approaches including term mining and training using manually extracted tracing matrices to cope with this issue and improve the precision of automated trace-retrieval methods.
Our work proposes another approach to address this terminology issue by introducing a novel process based on user interface labels.

Some traceability tools propose traceability links management such as TOOR developed by Pinheiro and Goguen \cite{pinheiro1996}. Those tools do not include traceability link recovery features. Asuncion et al. \cite{asuncion2006} provide a software application intended to support traceability link recovery. However, this software application is based on automated prospective traceability and does not use information-retrieval methods.

De Lucia et al. \cite{lucia2007recovering} developed ADAMS, a traceability link recovery tool based on latent semantic indexing to retrieve traceability links. Leuser and Ott use a tool called TraceTool \cite{leuser2010} implementing two IR algorithms (tf/idf and LSI) to evaluate various optimizations on large specifications written in German. Klock et al. generalized this approach with Traceclipse, an Eclipse plug-in providing a generic platform for integrating traceability link recovery into an IDE \cite{klock2011traceclipse}. Those authors do not explore the terminological richness of user-interface labels.

\section{Conclusion}
\label{sec:conclusion}
In this paper, we have introduced a new approach for traceability link recovery using the user interface texts. 
This approach leverages the terminological richness of UI labels in order to recommend relevant code elements.
Our prototype system uses a variety of optimizations related to indexing, relations between UI labels and code, etc.
We have evaluated our approach using a commercial system that implements a European regulation in the domain of green building.
A domain expert has assessed 1000+ recovered traceability links. The evaluation concludes that our system is very often able to provide a relevant code item in the top-5 results.
An area of future work would be to compare our approach based on user-interfaces against one based on code symbols. We are also trying to get an industrial impact with a free plugin implementing the traceability approach in the IntelliJ integrated development environment.

\bibliographystyle{plain}
\bibliography{references}

\begin{thebibliography}{10}

\bibitem{LUCENE}
Lucene homepage.
\newblock \url{http://lucene.apache.org/core/}.

\bibitem{SPOON}
Spoon homepage.
\newblock \url{http://spoon.gforge.inria.fr}.

\bibitem{AFNOR2001}
AFNOR.
\newblock Thermal performance of buildings - determination of air permeability
  of buildings - fan pressurization method.
\newblock Technical Report NF-EN-13829, AFNOR, 2001.

\bibitem{antoniol2002recovering}
Giuliano Antoniol, Gerardo Canfora, Gerardo Casazza, Andrea De~Lucia, and
  Ettore Merlo.
\newblock Recovering traceability links between code and documentation.
\newblock {\em Software Engineering, IEEE Transactions on}, 28(10):970--983,
  2002.

\bibitem{asuncion2006}
Hazeline Asuncion, Fr{\'e}d{\'e}ric Fran{\c c}ois, Richard~N. Taylor, Hazeline
  Asuncion, Fr{\'e}d{\'e}ric Fran{\c c}ois, and Richard~N. Taylor.
\newblock An end-to-end software traceability tool in an industrial context,
  2006.

\bibitem{cleland2010machine}
Jane Cleland-Huang, Adam Czauderna, Marek Gibiec, and John Emenecker.
\newblock A machine learning approach for tracing regulatory codes to product
  specific requirements.
\newblock In {\em Proceedings of the 32nd ACM/IEEE International Conference on
  Software Engineering-Volume 1}, pages 155--164. ACM, 2010.

\bibitem{hayes2006advancing}
Jane~Huffman Hayes, Alex Dekhtyar, and Senthil~Karthikeyan Sundaram.
\newblock Advancing candidate link generation for requirements tracing: The
  study of methods.
\newblock {\em Software Engineering, IEEE Transactions on}, 32(1):4--19, 2006.

\bibitem{klock2011traceclipse}
Samuel Klock, Malcom Gethers, Bogdan Dit, and Denys Poshyvanyk.
\newblock Traceclipse: an eclipse plug-in for traceability link recovery and
  management.
\newblock In {\em Proceedings of the 6th international workshop on traceability
  in emerging forms of software engineering}, pages 24--30. ACM, 2011.

\bibitem{leuser2010}
J{\"o}rg Leuser and Daniel Ott.
\newblock {\em Tackling Semi-automatic Trace Recovery for Large
  Specifications}, volume 6182, pages 203--217.
\newblock Springer Berlin Heidelberg, 2010.

\bibitem{lucia2007recovering}
Andrea~De Lucia, Fausto Fasano, Rocco Oliveto, and Genoveffa Tortora.
\newblock Recovering traceability links in software artifact management systems
  using information retrieval methods.
\newblock {\em ACM Transactions on Software Engineering and Methodology
  (TOSEM)}, 16(4):13, 2007.

\bibitem{marcus2003recovering}
Andrian Marcus and Jonathan~I Maletic.
\newblock Recovering documentation-to-source-code traceability links using
  latent semantic indexing.
\newblock In {\em Proceedings of the 25th International Conference on Software
  Engineering}, pages 125--135. IEEE, 2003.

\bibitem{NLCSPLUGIN}
Martin Monperrus and Antoine Mischler.
\newblock Natural language code search plugin for intellijidea.
\newblock \url{http://plugins.jetbrains.com/plugin/7288}, 2013.

\bibitem{pinheiro1996}
F.A.C. Pinheiro and Joseph~A. Goguen.
\newblock An object-oriented tool for tracing requirements.
\newblock {\em Software, IEEE}, 13(2):52--64, 1996.

\bibitem{sannier:tel-00941881}
Nicolas Sannier.
\newblock {\em {INCREMENT une approche hybride pour mod{\'e}liser et analyser
  dans le large les exigences r{\'e}glementaires de s{\^u}ret{\'e}}}.
\newblock PhD thesis, Universit{\'e} Rennes 1, December 2013.

\bibitem{spanoudakis2005software}
George Spanoudakis and Andrea Zisman.
\newblock Software traceability: a roadmap.
\newblock {\em Handbook of Software Engineering and Knowledge Engineering},
  3:395--428, 2005.

\bibitem{winkler2010survey}
Stefan Winkler and Jens Pilgrim.
\newblock A survey of traceability in requirements engineering and model-driven
  development.
\newblock {\em Software and Systems Modeling (SoSyM)}, 9(4):529--565, 2010.

\end{thebibliography}

\end{document}